\documentclass{osa-article}

\journal{oe}

\articletype{Research Article}

\usepackage{lineno}

\begin{document}

\title{Correcting photodetector nonlinearity in dual-comb interferometry}

\author{Philippe Guay,\authormark{1} Nicolas Bourbeau H\'{e}bert,\authormark{1,2} Alex~Tourigny-Plante,\authormark{1} Vincent Michaud-Belleau,\authormark{1,3}  and J\'{e}r\^{o}me Genest\authormark{1}}

\address{\authormark{1}Centre d'optique, photonique et laser, Universit\'{e} Laval, Qu\'{e}bec, Qu\'{e}bec G1V 0A6, Canada\\
\authormark{2}Currently with Institute for Photonics and Advanced Sensing (IPAS) and School of Physical Sciences, University of Adelaide, Adelaide, SA 5005, Australia\\
\authormark{3}Currently with LR Tech inc., L\'{e}vis, QC G6W 1H6, Canada}

\email{\authormark{*}philippe.guay.4@ulaval.ca} 



\begin{abstract}
Photodetector nonlinearity, the main limiting factor in terms of optical power in the detection chain, is corrected to improve the signal-to-noise ratio of a short-time measurement in dual-comb spectroscopy. An iterative correction algorithm minimizing out-of-band spectral artifacts based on nonlinearity correction methods used in classical Fourier-transform spectrometers is presented. The exactitude of the nonlinearity correction is validated using a low power linear measurement. Spectroscopic lines of H$^{12}$CN are provided and the error caused by the saturation of the detector is corrected yielding residuals limited by the measurement noise. 
\end{abstract}

\section{Introduction}
High precision spectroscopy measurements such as molecular identification and quantification \cite{GUA18,POL15} have high requirements in terms of signal-to-noise ratio (SNR). Optical frequency combs provide high brightness sources that can be used to yield high spectral resolution in configurations such as dual-comb spectroscopy (DCS). Despite the availability of significant optical power, usable power levels are almost always limited by nonlinearity (NL) in the detection chain \cite{ROY12,GUA19,HEB17,BAU11,ZOL13}.  This forces longer measurement durations to meet SNR requirements. This comes at a high cost since SNR improves only with the square root of measurement time. Furthermore, compensating low signal with increased measurement time is not always possible for several short-lived time-dependant phenomena such as rapid molecular detection \cite{THO06,SCH05}, combustion analysis, explosive reactions \cite{HIL09} and breath analysis \cite{NAR01,KHA01}.    

Increasing detected power levels is an efficient way to increase SNR, as improvements can scale linearly with the signal. This approach is however quickly limited by photodetector (PD) linearity. When detector saturation is reached, the voltage produced by the PD displays a nonlinear relation with the input power. The measured signal is consequently distorted by nonlinearity.

Commercial detectors from Thorlabs' PDB series are frequently used to measure short pulses  \cite{LU12,SOR21,TU15,ZHA16,GUA19,FDI19,HEB17,ZHAO16,CHE19,KIM14,BEN12,PEL14,DIA14,CHE15,SHI18,HUA17,EOM08}, but it has been shown that the impulse response displays nonlinear characteristics well below their specified continuous power saturation level \cite{GUA21a}. In the common instances where these detectors are used with short pulses, spectral nonlinear artifacts \cite{LAC00} arise and negatively impact the signal. For DCS experiments, the measured nonlinear interferogram  $(\text{IGM}_{\text{NL}})$ can be written as

\begin{equation}\label{poly}
\text{IGM}_{\text{NL}} =  a_0 + a_1[\text{IGM}_{\text{L}}]  + a_2[\text{IGM}_{\text{L}}]^2  + ...
\end{equation}

where  $\text{IGM}_{\text{L}}$ is the linear interferogram and where $a_0$, $a_1$ and $a_2$ are respectively the constant coefficient, the linear coefficient and the NL coefficient of second order. In the spectral domain, the n-th nonlinearity order is understood as the (n-1)-times convolution of the spectrum with itself. In occurrence, the second order nonlinear term is the auto-convolution of the spectrum such that it has a contribution at twice the frequency of the signal (2$f$) and also around DC. The third order term has a contribution at three times the frequency (3$f$) of the linear signal, but also generates content overlapping at $1f$. These spectral artifacts distort the signal in such a way that gas absorption lines strength are overestimated in transmittance measurement. For a complete and detailed description of the nonlinearity model, the reader is referred to \cite{GUA21b}. 

In this paper, a nonlinearity correction algorithm applicable to dual-comb spectroscopy is presented. It is based on minimizing the out-of-band spectral artifacts, a technique originally developed for Michelson-based Fourier-transform spectrometers \cite{LAC00,CHA84,GUE86,CAR90,ABR94,JES98}.  The validity of the correction process is assessed by comparing H$^{12}$CN spectroscopic lines with a time-averaged low power measurement for which NL impact is negligible. As a result, the absorption lines distorted by nonlinearity are corrected and thus match the low power measurement, but with a better SNR and for a significantly shorter measurement.

\section{Experimental methods}

\begin{figure}[htbp]
\centering\includegraphics[width=8cm]{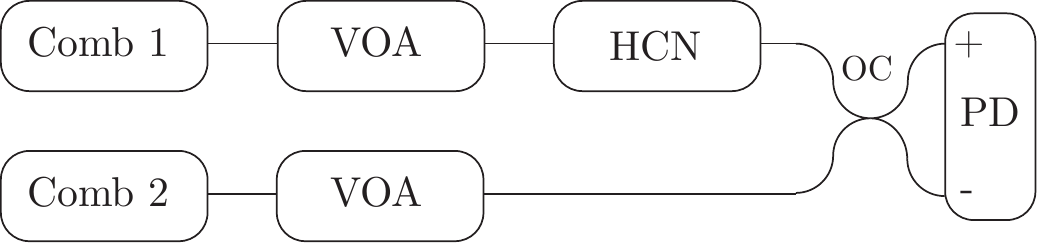}
\caption{\label{fig:expsetup} Experimental setup as a simplified block diagram. VOA: Variable optical attenuator. HCN: gas cell. OC: 50/50 output coupler. PD: Photodetector.} 
\end{figure}

The experimental setup is shown in Fig. \ref{fig:expsetup} in which two mode-locked lasers based on semiconductor saturable absorber having a repetition rate of 160 MHz were used \cite{SIN15}. These frequency combs centered at 1550 nm were sent through an optical coupler and then to a balanced amplified photodetector (Thorlabs PDB480C). The power sent on the detector was adjusted using variable optical attenuators following the combs. The lasers repetition rates were tuned so that the repetition rate difference ($\Delta f_r$) is about 150~Hz. The carrier-offset frequency of the lasers ($f_{\text{CEO}}$) was also adjusted so that the IGM's spectrum is centered at 20~MHz. A H$^{12}$CN acetylene cyanide gas cell was inserted in the setup to add spectral features. As nonlinearity generates spectral artifacts that may overlap the signal's spectrum, known spectral features such as absorption lines allow evaluating the impact of nonlinearity. Moreover, the retrieval of true line intensities enables assessing the validity of the nonlinearity correction. 

Low and high power IGMs were acquired, aligned, phase-corrected \cite{HEB17,GUA18}, and averaged. The dual-comb system used was minimally stabilized with a carrier offset envelope frequency lock and an optical lock  referenced about 20 nm away (1564 nm) from the absorption lines. Phase correcting the combs at 1564 nm usually leaves small repetition rate drifts that are corrected with a self-correction algorithm \cite{HEB19} that requires a continuous train of IGMs, but these drifts were not corrected here. Only the central portion of the IGMs was digitized (10\% of the IGM's length leading to a $\approx$1.6 GHz bin spacing) to reduce memory needs and computation time. While the truncation of IGMs maintains a sufficient resolution to observe narrow spectral features such as absorption lines, phase correcting the IGMs based solely on the phase at the zero path difference (ZPD) introduces an instrument lineshape that degrades the resolution \cite{POT13}. As a result, drifts in the lasers' repetition rates precludes the comparison with line-resolved experiments or with the HITRAN database. However, since the results are based on relative differences between absorption features in the linear, nonlinear, and NL corrected IGMs, this does not not affect the conclusions.

\begin{figure}[htbp]
\centering\includegraphics[width=8cm]{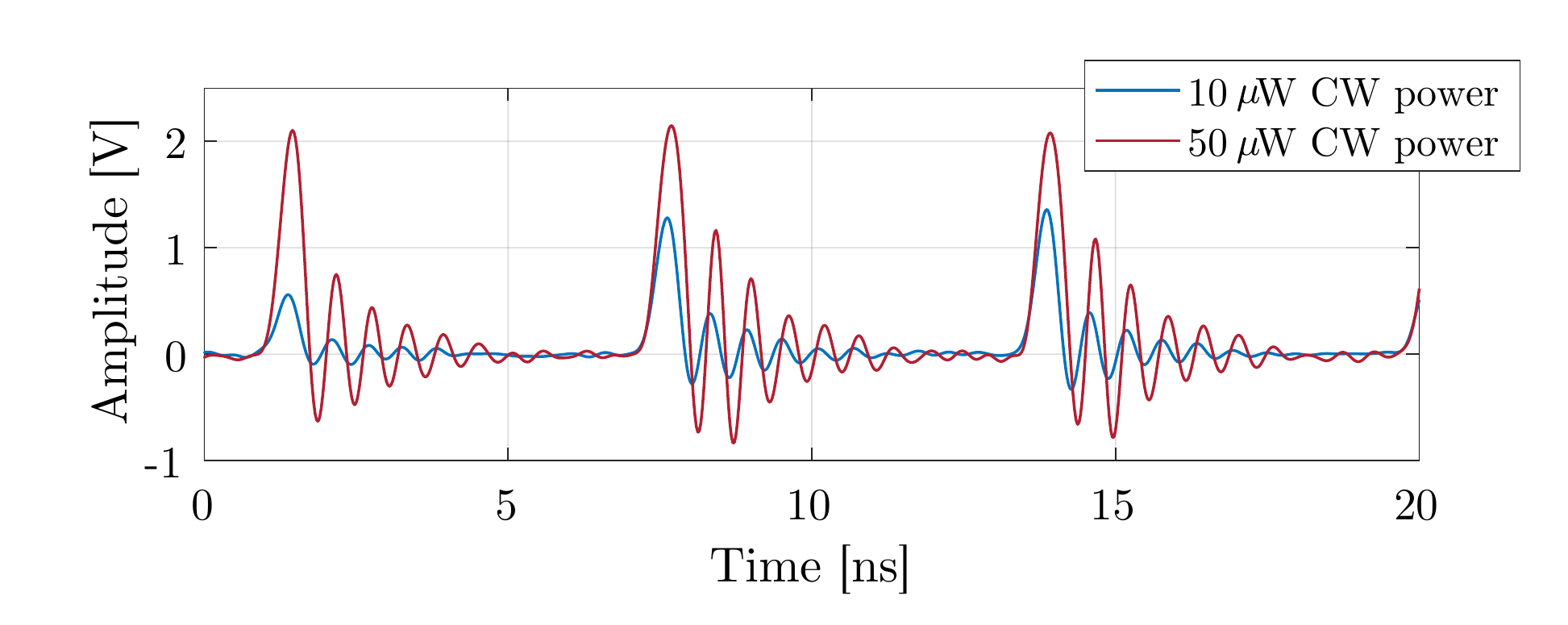}
\caption{\label{fig:sep_imp_resp} Detector's impulse responses at the interferogram's centerburst showing separate impulse responses.}
\end{figure}

In this experiment, the balanced photodetector PDB480C was chosen for its high bandwidth (1.6 GHz) to minimize the dynamic impact of nonlinearity. The bandwidth of the detector is ten times the repetition rate of the lasers, allowing to properly observe distinct impulse responses each time an optical pulse hits the photodiode as can be seen in Fig.\ref{fig:sep_imp_resp}. This aspect here is critical since the impulse responses of the detector have shown to broaden and ring significantly at high incident powers \cite{GUA21a}. Each overlapped successive impulse response whose shape is power-dependent has a contribution from the previous pulse that may not be proportional to it, making it impossible to determine the incident power on the detector. Here, since the impulse responses remain separated even for the most saturated samples measured, this dynamic NL effect is negligible and the static nonlinearity model presented in \cite{GUA21b} remains valid. 

\section{Correction algorithm}

\begin{figure}[htbp]
\centering\includegraphics[width=10cm]{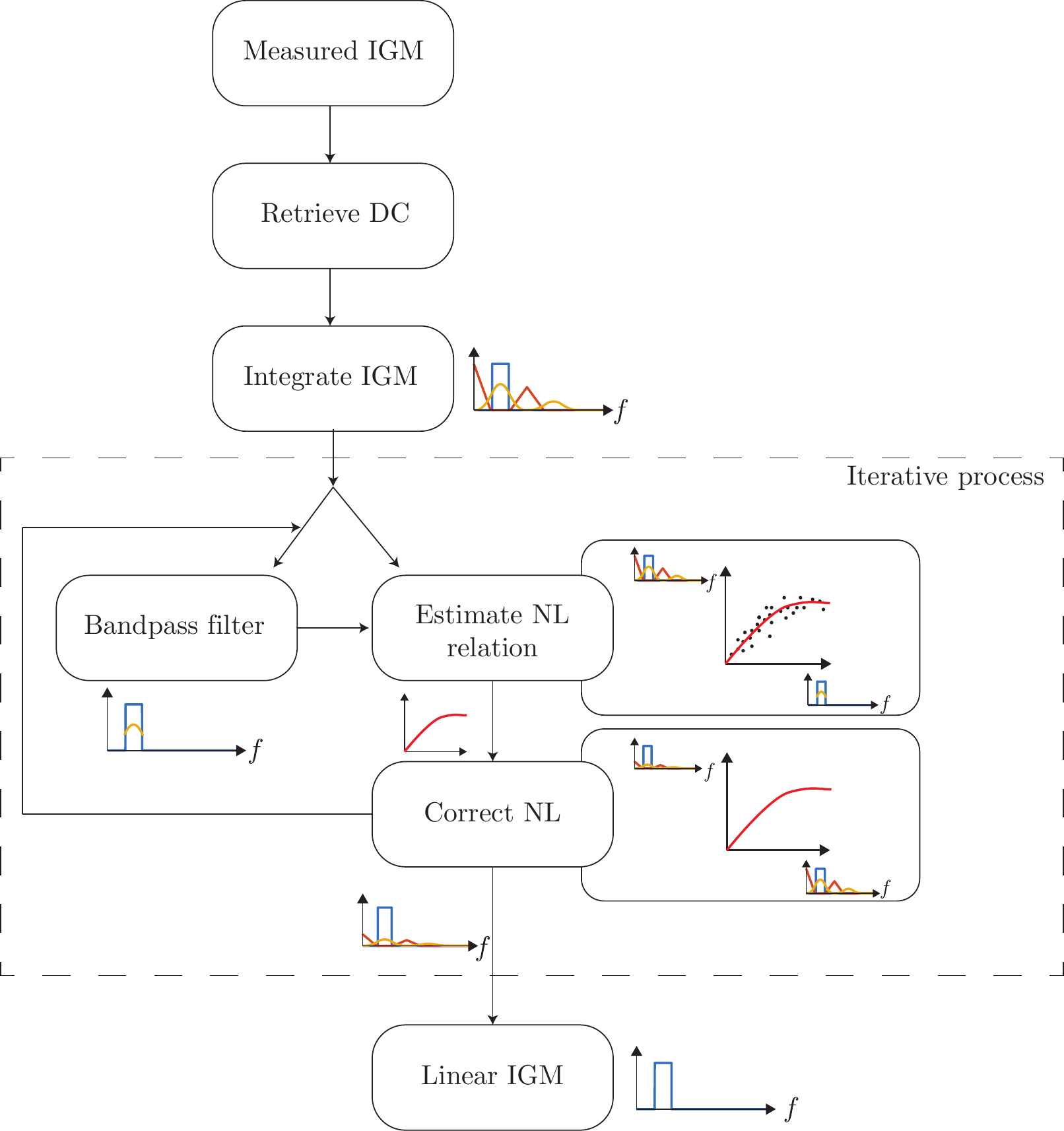}
\caption{\label{fig:algo} Flowchart of the nonlinearity correction process.}
\end{figure}

The NL correction procedure shown in Fig. \ref{fig:algo} is described here. First, since the detector is AC-coupled, the IGM needs to be AC-corrected. This means that the low frequency content around DC filtered out by the detector's high-pass filter is retrieved and added back to the IGM. This can later help the NL correction since artifacts of even orders are present around DC.

The AC-correction involves a nonlinear filter. Conceptually, this can be achieved by looking at the signal level between the pulses. Since there is no incoming power at these points, any observed non-zero signal must arise from the high pass filter compensating for the pulses areas in order to output a zero mean signal, averaged over several pulse periods. As a result, the signal between pulses is representative of what needs to be restored. In practice however doing so leads to a poor estimation of the content around DC. Another method takes advantage of the pulsed nature of the signal to extract information from the periodic spectrum. Since all spectral aliases contain very similar information, one can simply retrieve the DC content using a band-pass filter around the content at the pulse repetition frequency ($f_r$), perform an amplitude demodulation and restore the DC content with this information. In Fourier transform spectroscopy where signals are not pulsed and AC-coupled detectors are used, the DC content can be retrieved with the maximum amplitude of the signal at ZPD for a given modulation efficiency or through the integration of the spectrum \cite{LAC00}. In those instances, the AC-coupling frequency is typically orders of magnitude lower than in DCS, leaving the DC NL artifact mostly unaffected.

\begin{figure}[htbp]
\centering\includegraphics[width=10cm]{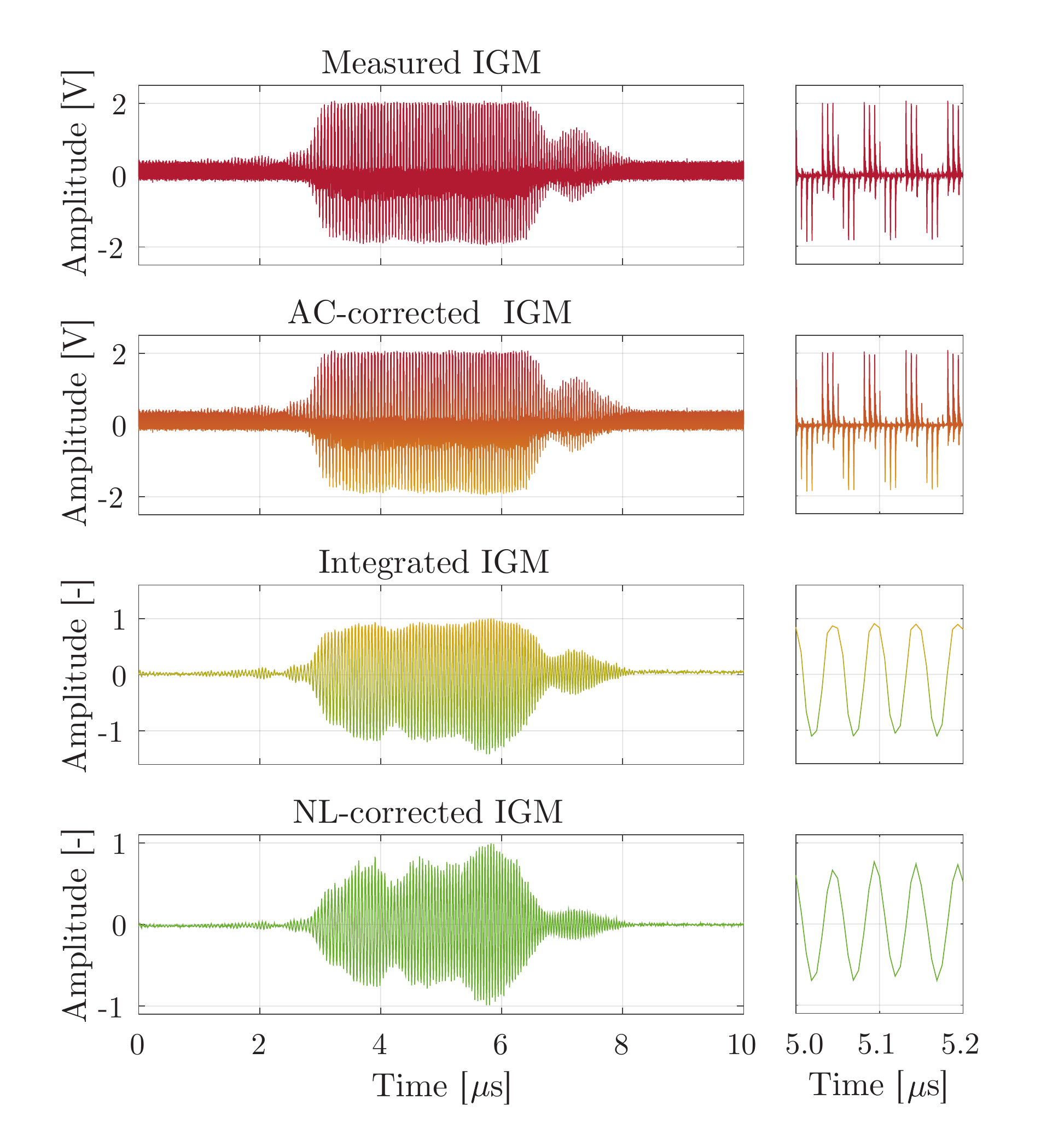}
\caption{\label{fig:IGMs} High power 50 $\mu$W measured interferogram (top panel), AC-corrected IGM (mid-top panel), IGM formed by the integrated pulses (mid-bottom panel), nonlinearity corrected IGM (bottom panel) }
\end{figure}

The measured high power nonlinear IGM is shown in the top panel of Fig. \ref{fig:IGMs}. The power sent to the detector is clearly saturating the output as the voltage rails at $\pm$~2V around ZPD. The AC-corrected IGM is shown in the middle-top panel of Fig. \ref{fig:IGMs}, where the differences with the measured IGM are very subtle. The AC-correction has not shown to have a strong impact since the following steps tend to minimize the spectral artifacts at DC. It nevertheless helps by bringing consistency to the 2nd order harmonics at 0 ad $2f$ and could prove critical if the artifact at $2f$ is not observable.

The second step of the correction algorithm consists in integrating the pulses. It has been previously demonstrated in \cite{GUA21a} that the area of the detector's impulse response is a more accurate indicator of the power sent on the photodetector than its amplitude. It was indeed observed that while the detector's amplifiers saturate, its impulse response broadens. Integrating the pulses is very similar to lowpass filtering to keep only the first alias of the measurement and downsampling the signal. The differences lie in the weighting function used to filter the signal. Here, the function chosen is simply a square box whose width is about one over the repetition rate of the lasers to avoid overlapping pulses that have different NL responses and thus mix NL and dynamic effects.

In theory, integrating the pulses rather than directly applying a low-pass filter also allows for an easier removal of timing jitter due to nonlinearity. As impulse response width varies with incoming power, the effective pulse location deviates from an equidistant $1/f_r$ grid depending on pulse amplitude. Assigning each integrated area to an exact time grid removes NL-induced sampling jitter, an effect that could otherwise induce unaddressed dynamic nonlinearity. 

In practice, this effect is present in our data and can be observed, but its impact is negligible in the overall NL correction method such that low-pass filtering the interferogram with a properly designed delay-less numerical filter having a $f_r/2$ cutoff frequency produces very little difference in the corrected interferogram. In the worst case observed, pulses have shown to shift by approximately 5\% of the pulse period. The integrated IGM shown on the middle-bottom panel of Fig. \ref{fig:IGMs} is now decimated as the integration of the pulses has downsampled the signal.



The integrated IGM is then corrected for nonlinearity using an iterative process that finds the static nonlinear response by minimizing out-of-band spectral artifacts. These operations are enclosed within the dashed square on Fig. \ref{fig:algo}. The nonlinear IGM is filtered at $1f$, which is used to obtain a first estimate of the linear signal thereafter referred to as the linear estimate. The filter pass and stop bands can be adjusted to choose the spectral content that is deemed valid and the frequency ranges over which spectral content shall be minimized. In practice, the expected spectral range of the linear signal is conserved and the spectral locations of the 2nd and 3rd NL artifacts at $2f$ and $3f$ are minimized. The range around DC is also usually minimized since all even NL harmonics generate content in that region.    

Notice that this necessarily requires a non-zero spectral width where spectral artifacts are not overlapped with the spectral content at $1f$. In DCS measurements, this implies that some of the available bandwidth between 0 and $f_r/2$ must be sacrificed to observe non-overlapped spectral harmonics. This means that the combs' repetition rate difference must be carefully adjusted and is in general smaller than theoretically possible when considering only spectral aliasing of the $1f$ frequency content. However, allowing more power on the detector is generally favorable over larger measurement bandwidth. 

The linear estimate is not the proper linear IGM since all NL harmonics take away energy from this band to generate artifacts at $2f$, $3f$, ... Furthermore, all odd harmonics and also aliased ones create overlapping spectral content around $1f$. This is why the nonlinear relation is estimated iteratively by computing progressively better estimates of the linear signal.

The linear estimate is used to estimate a static nonlinear correction curve. This relation between the measured NL IGM and the linear estimate is determined by a polynomial fit, where the polynomial terms correspond to nonlinearity of 2nd, 3rd order, etc. This fit describes the relation between the linear estimate (x-axis) and the nonlinear IGM (y-axis). Since the nonlinear IGM is the one measured, the relation has to be inverted to determine the linearized IGM.  A simpler practical approach however involves directly estimating the inverse relation: the measured nonlinear signal is instead put on the x-axis and the linear estimate on the y-axis. The resulting scatter plot can be fitted with a polynomial, the coefficients of which can be directly used to generate a linearized IGM from the measured NL IGM. This linearized IGM has reduced nonlinear artifacts and is fed back to the band-pass filter to iteratively obtain better estimates of the linear IGM until an exit condition is reached on the total estimate change since the last iteration. The flowchart in Fig. \ref{fig:algo} show that the nonlinear IGM is always used in the x-axis while the new linear estimate is replaced on the y-axis at each iteration. The resulting IGM is shown in the bottom panel of Fig. \ref{fig:IGMs}.

This procedure yields not only the corrected linear IGM but also the static nonlinear relation in the form of a high-order polynomial and can be used to correct subsequent data. Since the inverse of a low-order polynomial is not a low-order polynomial, it is necessary to estimate the inverse NL relation to higher orders. In practice a 10-th order polynomial is generally needed to properly correct up to 3rd order NL. Applying the NL correction procedure on a short ZPD-centered version of the IGM first and then correcting the full IGM is done here. It helps improve NL correction by feeding the algorithm only the portion of the IGM where NL is dominant.

\section{Results}

\begin{figure}[htbp]
\centering\includegraphics[width=10cm]{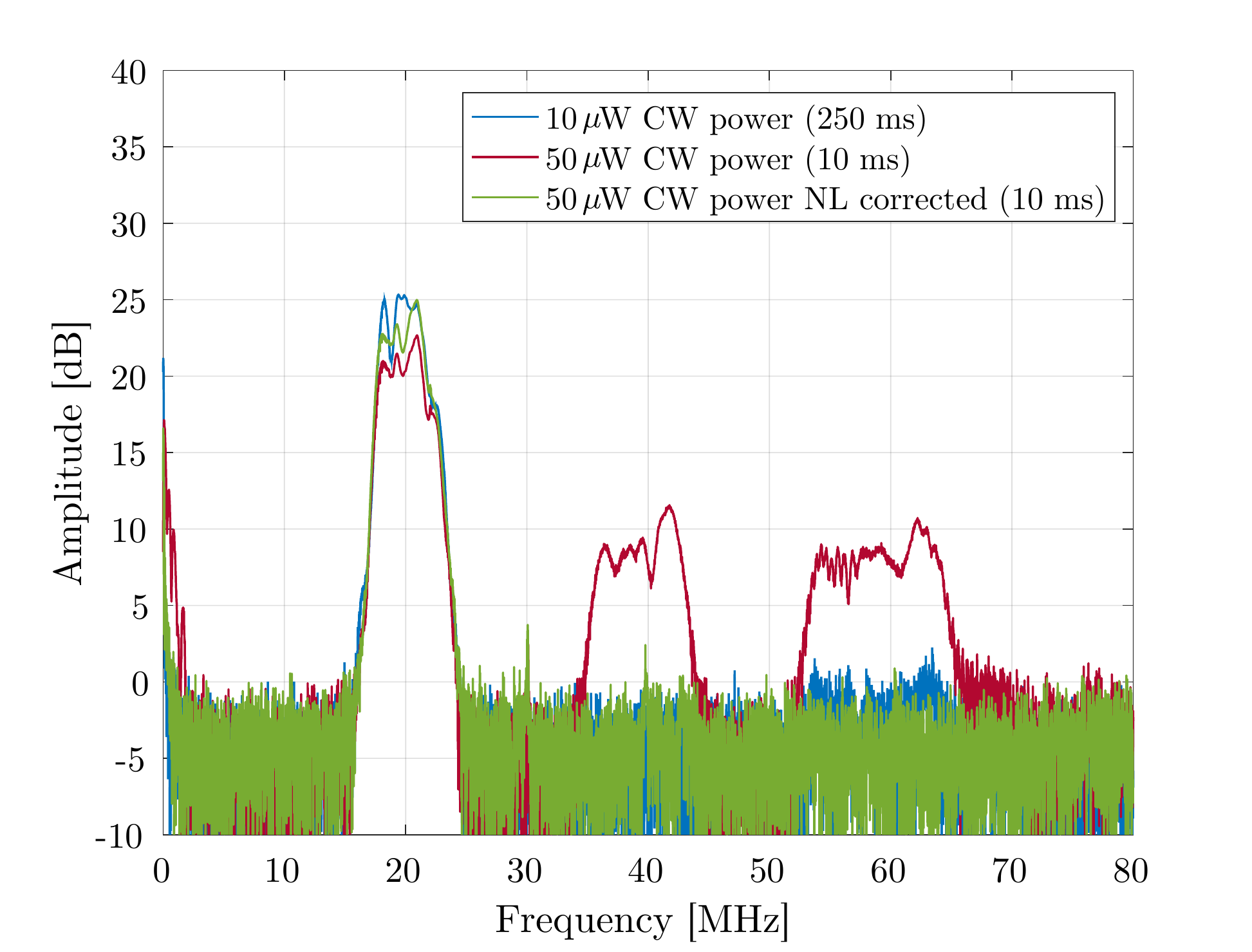}
\caption{\label{fig:spectrum} Low resolution spectra for the low power 10 $\mu$W (blue) and high power 50 $\mu$W before (red) and after (green) nonlinearity correction. Truncated IGMs have been averaged for the durations specified in the legend to have comparable SNR.}
\end{figure}

The spectra for a linear IGM measured with 10 $\mu$W of power as well as a nonlinear and corrected IGM measured with 50 $\mu$W are shown in Fig. \ref{fig:spectrum}. These spectra have been computed on short versions of the signals to allow better visualisation of NL artifacts. For the linear and the nonlinear measurements, the spectra are computed for a window of 100~$\mu$s centered on ZPD (IGM full length of 1/$\Delta fr$ = 6666 $\mu$s) and averaged for 2500 IGMs and 100 IGMs respectively. The second and third order spectral artifacts generated by the detector nonlinearity are clearly visible on the high power measurement (red) respectively at 40 MHz and 60 MHz. The fourth order may not have sufficient SNR to stand out of the noise, while the subsequent orders are folded in the $f_r/2$ band by optical sampling. For the low power IGM, spectral artifacts are barely visible, hence confirming that the low power measurement is essentially linear (NL 25 dB below signal) and that the measured absorption features are not significantly distorted.

The high power measurement has been corrected by the previously described process to yield the spectrum shown in green on Fig. \ref{fig:spectrum} which shows no spectral artifacts left at 40 MHz and 60 MHz. This suggests that the third order nonlinearity falling directly on the band of interest around 20 MHz has also been corrected and that the gain error \cite{GUA21b} due to second order nonlinearity has been corrected. In this case, the algorithm converged to 1~ppm in 29 iterations

The spectra of the NL corrected and linear IGMs (green and blue on Fig. \ref{fig:spectrum}) do not perfectly overlap each other at 20 MHz as one would expect. This is explained by polarization adjustments made to the setup in between measurements to ensure maximal modulation efficiency of the IGMs.

\begin{figure}[htbp]
\centering\includegraphics[width=10cm]{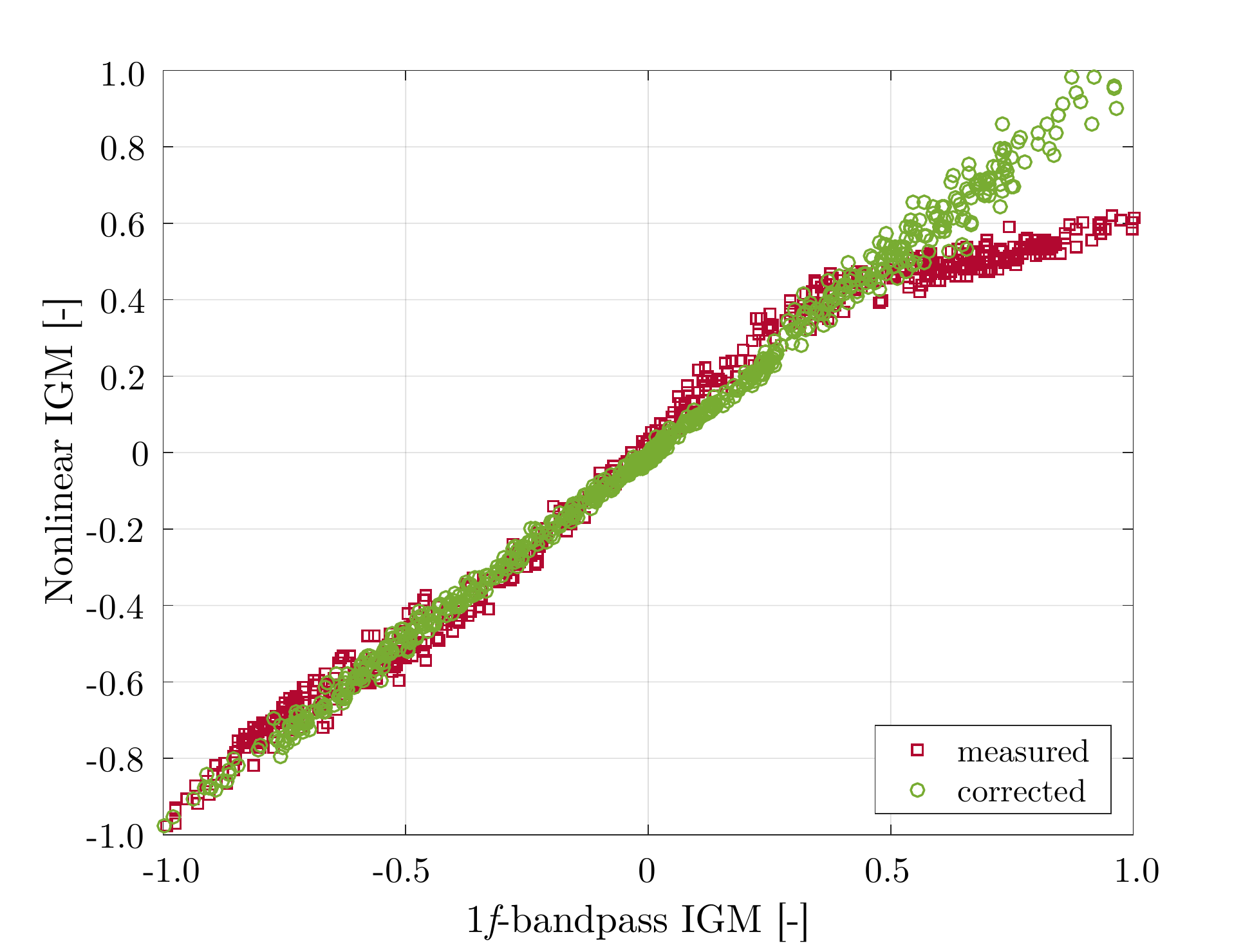}
\caption{\label{fig:8figure} Nonlinear IGM plotted against its $(1f)$ filtered version before (red squares) and after (green circles) nonlinearity correction }
\end{figure}

As an indicator of the correction efficiency, a nonlinear IGM (measured and corrected) is plotted in Fig. \ref{fig:8figure} against an IGM filtered in a 10 MHz bandwidth around 20 MHz ($1f$). A perfectly linear IGM and its $1f$-filtered version are expected to display a linear relation with unit slope as both signal are the same since there is no content outside the $1f$ band. It can be seen that the measured IGM is far from that instance. The saturation of the detector, mainly of the positive photodiode, causes a kink in the upper right part of the figure. As the amplitude of the $1f$-filtered IGM increases, the measured IGM reaches saturation. This is one effect that the NL correction algorithm aims to remove. As it can be seen from the corrected data, the curve straightens up. 

The noise trace on Fig. \ref{fig:8figure} becomes larger where the kink was located on the measured signal. This is caused by the gain that is applied to the saturated signal at ZPD. Since the measurement noise originating from the oscilloscope has the same variance across the entire duration of the IGM, the correction at ZPD where the IGM was saturated also acts as a gain on the noise. NL correction thus implies inducing noise nonstationarity and correlations can be expected in the noise power spectral density.

\begin{figure}[htbp]
\centering\includegraphics[width=10cm]{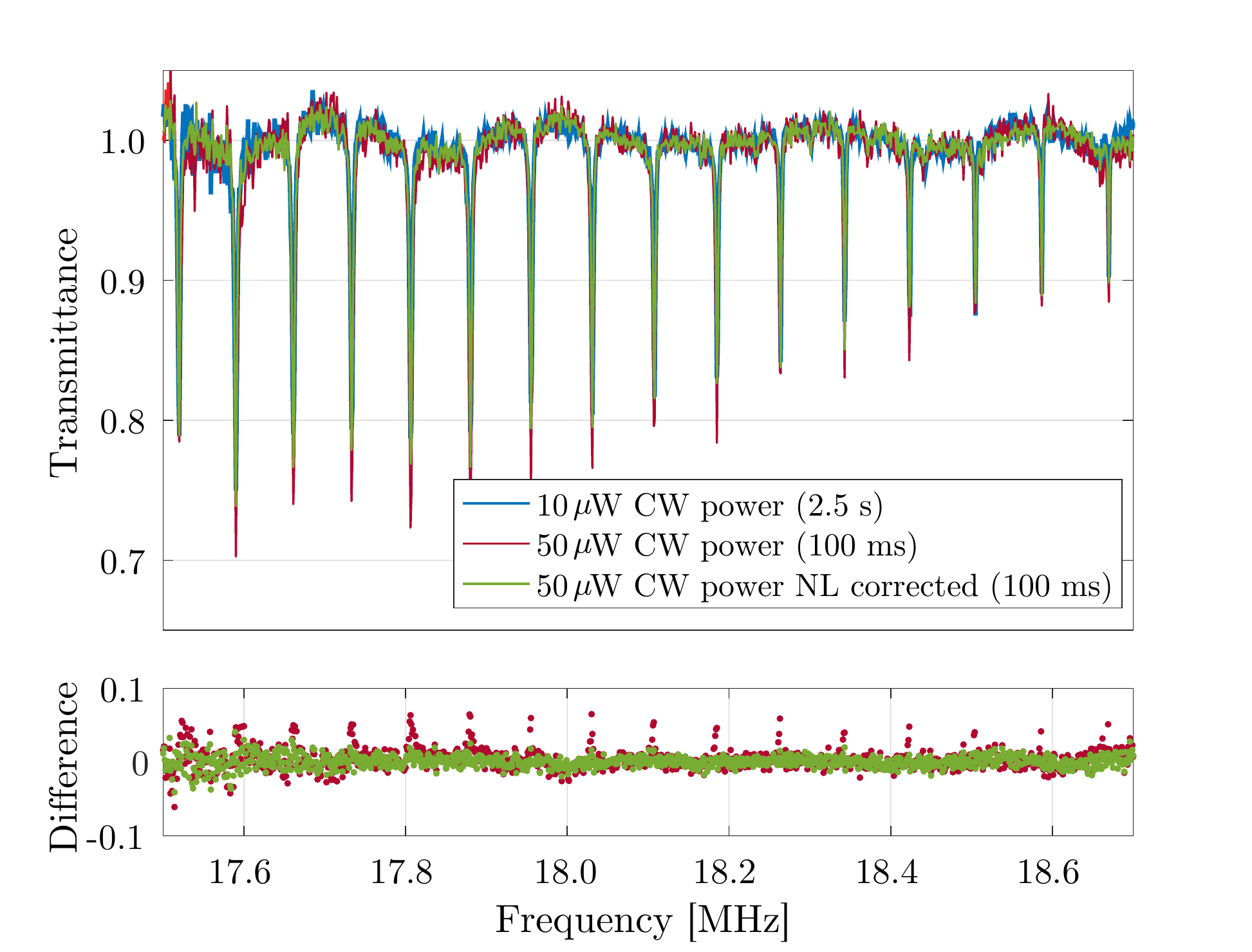}
\caption{\label{fig:spectro} (top panel) Absorption lines in the P branch of H$^{12}$CN for the low power (blue), high power before (red) and after (green) nonlinearity correction. (bottom panel) Differences between the high power (measured and corrected) and the reference low power measurements.}
\end{figure}

The dual-comb system is then used to measure the absorption lines of hydrogen cyanide H$^{12}$CN to assess the efficiency of the nonlinearity correction. Lines in the P branch are shown in Fig. \ref{fig:spectro} for the low power and high power measurements. Longer versions of the IGMs (1~ms window centered on ZPD for IGM full length of 1/$\Delta fr$ = 6.6 ms) have been used to provide sufficient resolution to observe absorption lines. For the low power measurement, 2500 IGMs of 1~ms were processed to reach an SNR equivalent to the 100 high power IGMs acquired for 1~ms. This validates that increasing the power sent on a PD by a factor of 5 allows reducing the measurement time by a factor of 25. The nonlinear measurement shows stronger absorption features with deepened lines, causing ripples where the lines are located in the difference between these measurements. For the NL corrected case, the ripples are significantly reduced and the difference becomes mainly dominated by oscilloscope measurement noise. This validates that the NL correction procedure corrects distortion in absorption features caused by the saturation of the photodetector. 
The slow varying trend in the transmittance baseline shown in the top panel of Fig. \ref{fig:spectro} is due to the lack of a proper calibration of this transmittance measurement. A single 8th degree polynomial was used to remove the baseline rather than a measurement with an empty gas cell.

\section{Conclusion}
As a conclusion, it was demonstrated that saturating the detector does not prevent reliable and precise spectroscopic measurements when the detector's bandwidth allows separate impulse responses and when nonlinearity is properly taken into account. To manage nonlinearity, a correction algorithm based on existing Fourier transform spectrometer methods was presented and the exactitude of this correction was validated with the measurement of H$^{12}$CN absorption lines.

\begin{backmatter}
\bmsection{Funding}
This work was supported by Natural Sciences and Engineering Research Council of Canada (NSERC), Fonds de Recherche du Qu\'{e}bec - Nature et Technologies (FRQNT) and the Office of Sponsored Research (OSR) at King Abdullah University of Science and Technology (KAUST) via the Competitive Research Grant (CRG) program with grant \# OSR-CRG2019-4046

\bmsection{Acknowledgments}
The authors thank Ian Coddington at NIST for providing the dual-comb system.

\bmsection{Disclosures}
The authors declare no conflicts of interest.

\bmsection{Data availability} Data underlying the results presented in this paper are not publicly available at this time but may be obtained from the authors upon reasonable request.

\end{backmatter}

\bibliography{sample.bib}

\end{document}